\documentstyle[psfig]{l-aa}
\begin{document}
\thesaurus{08         % A&A Section 6: Form. struct. and evolut. of stars
              (12.04.1;  % Dark matter
               12.07.1;  % Gravitational Lensing;
               08.12.2;  % Stars: ow-mass, brown dwarfs
               08.02.3;  % Binaries: general
               08.16.2)} % planetary systems

\title{Lensing of unresolved stars towards the Galactic Bulge.}
\author{C. Alard \inst{1,2}}
\institute
{DEMIRM,Observatoire de Paris, 61 Avenue de l'observatoire, F-75014
 Paris, France
 \and
Centre d'Analyse des Images de l'INSU, B\^atiment Perrault, 
 Observatoire de Paris, 61 Avenue de l'Observatoire, F-75014, Paris, France}
\offprints{C. Alard}
\date{Received ......; accepted ......}
\maketitle
\begin{abstract}
This paper presents an analysis of lensing of unresolved background stars.
Previous calculations of the lensing rates and optical depths considered only
 resolved stars. However, if a faint unresolved star lens is close enough
 to a resolved star, the event will be seen by the microlensing experiments
 and attributted to the bighter star.
 The blending biases the duration, making 
 the contribution of the unresolved stars very significant for
  short events. It is especially annoying, because this contribution is confused
 with lensing by brown dwarfs. The exact rates of these blended events are extremly sensitive to
 the limiting magnitude achieved in the microlensing search. 
 Appropriate calculations of the optical depth and rates are provided here,
 and illustrated in the case of the DUO and OGLE experiments. 
 The additional contribution of unresolved stars is very significant. 
 It probably 
 explains the high optical depth and rates observed towards the Galactic Bulge. 
 The blended unresolved event can be identified using either 
 the color shift or the light curve shape. 
 However,
 neither of these two methods is apropriate to identify a large number of blended events towards
 the Bulge. In some case of good photometry and small impact parameter an
  indentification is possible.
 This is illustrated by the case of the OGLE 5 event, which clearly appears as
 a case of lensing of an unresolved star.
 The recent results obtained by the PLANET collaboration indicate that a high resolution
 and dense sampling of the light curve is possible, and will probably provide a very interesting 
  possibility to correct the blending bias, as demonstrated for OGLE 5. This possiblity,
 is certainly better than a statistical estimation of the lensing rates, which are always prone to 
 some uncertainty. But, at this time, the analysis of Microlensing events  found in 
 the various microlensing experiments requires the uses of modelisations of the contribution of
 unresolved stars.  
\end{abstract}
\section{Introduction}
 In previous microlensing models the rates and optical depth per star are
 usually estimated (Kiraga and Paczy\'nski, 1994
 Han and Gould, 1995, Mollerach and  Roulet , 1996). 
 Below some limiting magnitude these stars can
 not be resolved individually and form a dense stellar background on all the surface of the 
 images. I will call these star "background stars". These background stars can be lensed, and this issue
 has already been investigated in the case of the LMC (Bouquet, 1993), and for the Galaxy (Nemiroff, 1994).
 However the stellar fields of the current
 Microlensing experiments MACHO (Alcock {\it et al.} 1996, Alcock {\it et al.} 1993), OGLE
 (Udalski{ \it et al.} 1994), 
 DUO (Alard 1995a, Alard {\it et al.} 1995b)
 , EROS (Aubourg {\it et al.} 1993.) are extremly crowded, so tha
 inside the resolution radius of a monitored stars, there are plenty of these unresolved 
 background stars. Thus if one of these  background stars is lensed, we will see a blended event.
 The amplification, duration and impact parameter will be biased by the blending. 
 We may now wonder if their contribution to the total number of events observed is significant,
 compared to those expected from the resolved stars.
 As the blending of a background star with a resolved star will be high 
 we expect a large reduction of the effective Einsten Radius (Di Stefano and Esin, 1995). 
 Consequently the observed number of events will be a competition between the decrease of the
 Einsten radius, and the increasing number of stars at larger magnitudes.
 Also the effective duration of the event will be much shorter than it is for to events on resolved
 stars, and consequently it may produce a tail of short events. These short events
 could be confused with low mass lens events, and especially lensing by brown dwarf, so that 
 a detailed study of these events is particulary important.
 Another problem will be to recognise these blended events. The blending of light
 produces a modified light curve which might be distinguished from the unblended light
 curve. However to make an unambiguous separation a good time sampling is required, and a very accurate
 photometry is required. In the 
 case of these short events, we expect that it will be difficult to achieve a sampling dense enough.
\section{The Bulge luminosity function}
 The study of lensing of unresolved stars is very sensitive to the shape of the
 adopted luminosity function. 
 In previous works (Kiraga and Pazcy\'nski, Roulet and Mollerach, Han and Gould),
 the number of stars at a given luminosity $L$ is expressed as: $dN=L^{-\alpha}$,
 with usually $\alpha \simeq 2$. A more realistic luminosity function can be derived
 from the Holtzman (Holtzman {\it et al.} 1993) observations of a field in Baade Window with the HST.
 The final Holtzman {\it et al.} luminosity function can be approximated by three continuous
 straight segments
 with different slopes. The first change in the slope occur at $V \simeq 20$
. With the extinction and distance to the Galactic Center adopted by Holtzman {\it et al.} 
 , it correponds to a value of $M_V=4.9$.
 . The Holtzman {\it et al.} study agrees well with Terndrup (Terndrup 1989) investigations of the Galactic Bulge
 which found a value  of $M_V=4.07$ for the Bulge turn-off at $b=-8$. In the Holtzman luminosity function the turn-off is situated 
 around $M_V\simeq 4.0$ also.
   The second change in the slope occur around $V=21.5$ and seeems to continue until the
  Holtzman {\it et al.} limit situated at $V \simeq 22.5$. A comparison with the Luminosity function
 of the globular clusters reviewed by Mould (Mould 1996) indicates that the shape of their luminosity function
 is almost the same in this range of magnitude. The Mould diagram allows also to a better determination
 of the last segment of the luminosity function. This straight segment seems to hold 
 until $M_V=11$ where a possible turnover in the luminosity function is observed. However $M_V=11$ is
 about 7 magnitudes fainter than the limiting magnitude of the current microlensing surveys, and
 such faint unresolved stars do not contribute to the optical depth. The duration of the
 events on stars blended with a star 7 magnitudes brighter would be so short that it is not 
observable. Consequently the exact behavior of the luminosity function in this region is not important
 for our study, and we will make the reasonable choice to cut the luminosity function at $M_V=11$.
 Looking now at the bright side of the luminosity function, we find that at 
 magnitudes brighter than the turnoff , the slope is very steep
 indicating that the number of stars drops rapidly. In this region the Bulge stars are essentially
 sub giants. The density rises again in the clump giant region, however these stars do not makes
 more than 5 percent of the stars in the current surveys. Consequently they have not a very significant
 influence on the total luminosity function. These stars should be treated separately (Alard 1996),
 and in the present study, we will ignore them. 
 This will put the bright end of the luminosity function at $M_V \simeq 3$ 
 in the Holtzman {\it et al.} diagram. The adopted value for the distance to the Galactic
 center in this study is 8.5 Kpc. 
 The final adopted luminosity function is shown in figure 1. 
\begin{figure}
\centerline{\psfig{angle=0,figure=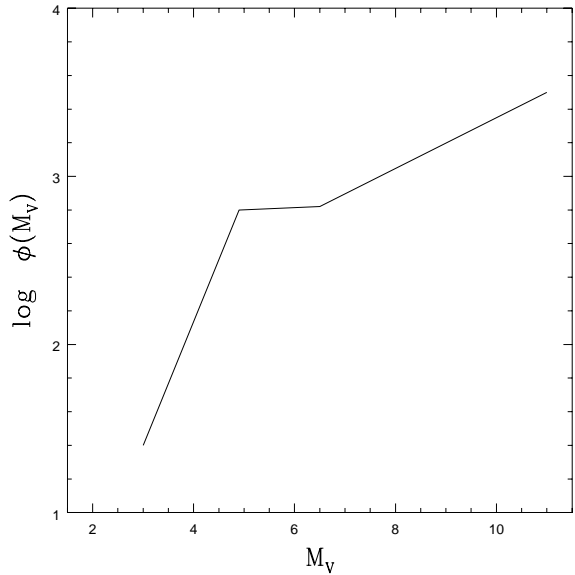,width=8cm}}
\caption{ The adopted Luminosity function for the present study.} 
\end{figure}
\section{Estimation of the lensing rates and optical depth}
 It is clear that, in an investigation of the lensing of unresolved stars,  consistency requires that
 we calculate the rates and optical depth also with the same luminosity function.
 This involves some slight changes from previous analysis. Therefore I 
 will first investigate the case of resolved stars considering a more general form
 for the luminosity function. I first calculate the change in the optical depth.
%
%%%%%%%%%%%%%%%%%%%%%%%%%%%%%%%%%%%%%%%%%%%%%%%%%%%%%%%%%%%%%%%%%%%%%%%%
                   \subsection{Optical depth}
%%%%%%%%%%%%%%%%%%%%%%%%%%%%%%%%%%%%%%%%%%%%%%%%%%%%%%%%%%%%%%%%%%%%%%%%
%
%

 The Numbers of source stars at distance $D_s$, and with absolute luminosity L
 can be written as:
 $$
 dN(D_s,L)= k \ lf(L)\ n(D_s) \ D_s^2 \ dL \ dD_s 
 $$
 Where k is a constant. \\
 Suppose that the experiment is able to find a total of $N_{tot}$ stars up to a 
 limit in apparent luminosity $l_0$. This leads to the following expressions:
  \begin{equation}
  \hspace{1cm} \frac{dN(D_s,L)}{N_{tot}}  = \frac{1}{I_d} \times lf(L)\ n(D_s) \ D_s^2 \ dL \ dD_s 
  \end{equation}
 with:
 $$
 I_d = {\int_{D_s} \int_{L=L_{min}}^{L=L_{max}} lf(L)\ n(D_s) \ D_s^2 \ dL \ dD_s}
 $$
and:
 $L_{min} = l_0 \times D_0^2/D_s^2$, where $l_0$ is the apparent limiting luminosity. \\
 $L_{max} = l_{sat} \times D_s^2/D_0^2$. 
 Where $l_{sat}$ is defined as the saturation limit of the detector. It can be expressed as: $l_{sat}=d \times l_0$,
 where $d$ is the detector dynamic range, a reasonable value is about 6 magnitudes in crowded fields, either for CCD or 
 photographic plates.
%
%%%%%%%%%%%%%%%%%%%%%%%%%%%%%%%%%%%%%%%%%%%%%%%%%%%%%%%%%%%%%%%%%%%%%%%%
%%%%%%%%%%%%%%%%%%%%%%%%%%%%%%%%%%%%%%%%%%%%%%%%%%%%%%%%%%%%%%%%%%%%%%%%

The optical depth associated with these stars can be calculated as (Paczy\'nski
 1991, Kiraga and Paczy\'nski 1994, hereafter refered as KP):
 \begin{equation}
 \hspace{1.5cm} d\tau_0(D_s,L)=\frac{dN(D_s,L)}{N_{tot}} \times f(D_s) 
  \end{equation}
with:
 $$
 f(D_s)=\frac{4 \pi G}{c^2} \int_0^{D_s} \rho_d(D_d)  \times \frac {D_d(D_s-D_d)}{D_s} \ dD_d
 $$
     and $l=L\times \frac{D_0^2}{D_s^2}$ \\
 To get the total optical depth we have now to integrate eq (1) over the source distances $D_s$
 and 
 the absolute luminosity $L$ of these stars.  \\
 $$
  \tau_0 =  \frac{1}{I_k} \int_{D_s} \ N_{L0}(D_s) \ f(D_s) \ n(D_s) \ D_s^{2-2 \beta} \ dD_s 
 $$
with:
$$
 N_{L0}(D_s)= \frac{I_k}{I_d} \times D_s^{2 \beta}  \int_{L=L_{min}}^{L=L_{max}}  \ lf(L) \ dL  
$$ 
and:
$$
I_k = {\int_{D_s} \ n(D_s) \ D_s^{ \ 2-2 \beta} \ dL \ dD_s}
$$
Here the factor $D_s^{2 \beta}$ is put in this expression to allow direct comparison
with the standard optical depth formulae (Kiraga and Paczy\'nski 1994): 
$$
  \tau_k=  \frac{1}{I_k}\int_{D_s} \ f(D_s) \ n(D_s) \ D_s^{ \ 2-2 \beta} \ dD_s 
$$
We see that the Kiraga and Paczy\'nski expression is simply equivalant to taking $N_{L0}(D_s)=1$
 at all distances. A slight improvement is to take $N_{L0}(D_s)=0$ beyond a given distance (Roulet
 and Mollerach, 1995).
 The plot in Figure 2 allows a direct comparison between $N_{L0}(D_s)$, and the approximation
 $N_{L0}(D_s)=1$. Figure 2 illustrates the calculations both for DUO and for the OGLE
 experiment, which is about one magnitude deeper. 
\begin{figure}
\centerline{\psfig{angle=0,figure=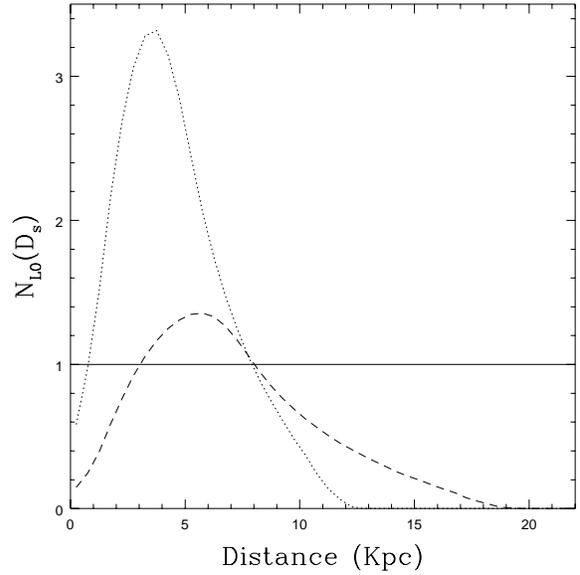,width=8cm}}
\caption{Plot of the $N_{L0}(D_s)$ function as a function of distance.
 For the DUO experiment a limiting magnitude of $M_V=3.5$ at the distance of the Galactic 
 Center was adopted, OGLE is assumed
 to be one magnitude deeper. The dotted line is for DUO, and the dashed line
 for OGLE.} 
\end{figure}
Let's see now the changes in the total optical depth introduced by the $N_{L0}(D_s)$ function.
The optical depths are computed using the COBE bar model for the Bulge (Dwek {\it et al.}, ..), and
 the Bahcall model for the Disk (Bahcall and Soneira, 1984)
The ratio of the modified optical depth to the standard one is shown in table 1. The change
is already important for OGLE, and is more significant again for DUO. It is easy to understand that 
the lower optical depth of the new formulae is essentially due to the fact that due to the limiting
 magnitude, the experiments are more sensitive to sources in the near end of the bar, for which 
 the density of lenses on the line of sight is smaller.
\begin{table} 
   \caption{Optical depth ratio for DUO and OGLE}
   \label{}
   \begin{tabular}{llll}
      \hline\noalign{\smallskip}
   Optical depth ratio & Bulge lenses & Disk lenses & Total \\ 
   \hline\noalign{\smallskip}
     OGLE & 0.85 & 0.89 & 0.86 \\ 
     DUO & 0.75 & 0.8 & 0.77 \\  
   \hline\noalign{\smallskip}
      \noalign{\smallskip} 
  \end{tabular}
\end{table}
%
%%%%%%%%%%%%%%%%%%%%%%%%%%%%%%%%%%%%%%%%%%%%%%%%%%%%%%%%%%%%%%%%%%%%%%%%
                   \subsection{Event rates.}
%%%%%%%%%%%%%%%%%%%%%%%%%%%%%%%%%%%%%%%%%%%%%%%%%%%%%%%%%%%%%%%%%%%%%%%%
%
The Formulae given for the lensing rates by Kiraga and Paczy\'nski is:
$$
 \Gamma_k =  \frac{1}{I_k} \int_{D_s} \ f_{\gamma}(D_s) \ n(D_s) \ D_s^{ \ 2-2 \beta} \ dD_s 
$$
Where:
$$
f_{\gamma}(D_s) = \frac{4 \ G^{1/2}}{c M^{1/2}} \int_0^{D_s}  \int_{V_i} 
 \rho_d(D_d) \ V(D_s,D_d) 
$$
$$
\times {[\frac {D_d(D_s-D_d)}{D_s}]}^{1/2} \ dD_d \ dV_i
$$
$V(D_s,D_d)$ is the relative velocity between lens and source, and the $V_i$ are
 the four components of this velocity. \\
Now, it is straightforward to introduce the $N_{L0}(D_s)$ function in this expression, exactly
 as was done for the optical depth. \\ \\
This leads to:
$$
 \Gamma_0 =  \frac{1}{I_k} \int_{D_s} \ N_{L0}(D_s) \ f_{\gamma}(D_s) \ n(D_s) \ D_s^{2-2 \beta} \ dD_s 
$$
The result of the calculation of the $\Gamma_0$ and $\Gamma_k$ rates is shown in figure 3 for
 DUO, and in figure 4 for OGLE. For all the figures the velocity dispersion given by Han
 and Gould (Han and Gould 1995) was adopted. The mass function is a Salpeter mass function
 with a lower cut off at $0.08 M_{\sun}$ and an upper cut off at $1 M_{\sun}$.
\begin{figure}
\centerline{\psfig{angle=0,figure=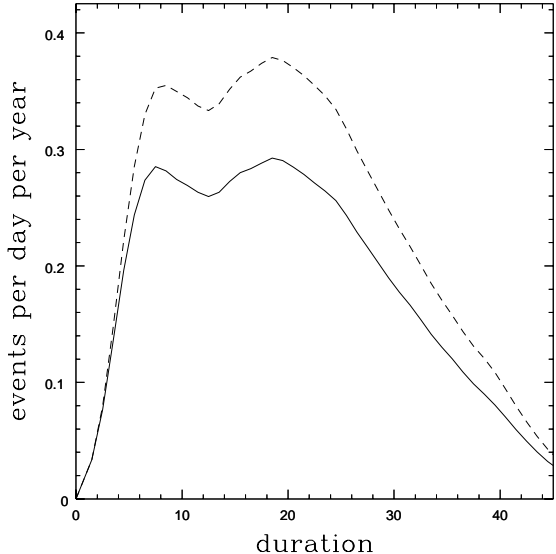,width=8cm}}
\caption{Comparison of the total rates of events for Bulge sources using
 KP formulae (dashed line), and the formulae described in this text (solid line). The total rate 
 is estimated for the 1994 season of the DUO experiment using the DUO efficiencies
 (Alard and Guibert 1996). The two bumps are due to the particular shape of
 the DUO efficiencies.} 
\end{figure}
\begin{figure}
\centerline{\psfig{angle=0,figure=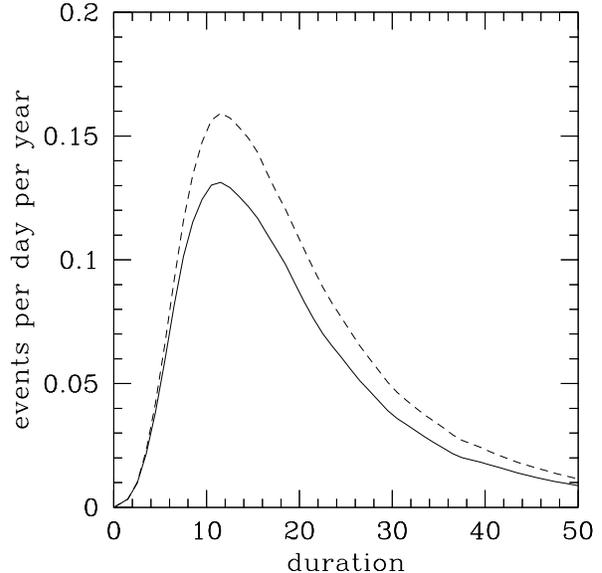,width=8cm}}
\caption{Comparison of the total rates of events for Bulge sources using
 KP formulae (dashed line), and the formulae described in this text (solid line). The total rate 
 is estimated for $10^6$ stars in Baade window using the OGLE efficiency
 (Udalsky {\it et al.} 1994)} 
\end{figure}
%
%
%%%%%%%%%%%%%%%%%%%%%%%%%%%%%%%%%%%%%%%%%%%%%%%%%%%%%%%%%%%%%%%%%%%%%%%%
%%%%%%%%%%%%%%%%%%%%%%%%%%%%%%%%%%%%%%%%%%%%%%%%%%%%%%%%%%%%%%%%%%%%%%%%
%
\section{Modelling the lensing of blended unresolved stars}
\subsection{Basic Principles}
The Einsten radius is defined as the distance for which the amplification of a microlensing event is
 a factor of 1.34.
In the case of an unresolved background star of luminosity $l_0$, blended with
a resolved star of magnitude $L_0$, the background star amplification required to
make a total amplification in the combined image of 1.34 is:
$$
 A_b = 0.34 \times f_b+1.34
$$
Hereafter, I will refer to $f_b$  as the blending factor:
$$
f_b=L_0/l_0
$$
This gives the following expression for the blended Einsten radius:
$$
 R_b(f_b)= R_e \times re_b(f_b)
$$
with:
$$
 re_b(f_b)= \left [2(-1+A_b/\sqrt{(A_b^2-1.0)} \right ]^{1/2}
$$
%
%%%%%%%%%%%%%%%%%%%%%%%%%%%%%%%%%%%%%%%%%%%%%%%%%%%%%%%%%%%%%%%%%%%%%%%%
       \subsection{Optical Depth due to the Unresolved stars.}
%%%%%%%%%%%%%%%%%%%%%%%%%%%%%%%%%%%%%%%%%%%%%%%%%%%%%%%%%%%%%%%%%%%%%%%%
%
 Let us consider unresolved background stars with absolute luminosity L at a distance $D_s$,
 whose number is dN($D_s$,L), and let us suppose also that they are blended with stars of 
 apparent luminosity $l_b$. The optical depth per resolved star associated with these stars can be calculated using
 equation (2) by
 replacing the Einsten radius by the blended Einsten radius. 
 $$
 d\tau(D_s,L)=\frac{dN(D_s,L)}{N_{tot}} \times r_b(l_b,l) \times f(D_s) \times R
 $$
 The expression for $d\tau$ contains the factor $R$, which takes into account that only
 those unresolved objects which are lensed close to a resolved star will be seen. The Value
 of $R$ will be close to the ratio of the area of the seeing disk, to the mean space 
 occupied by a star on the image.\\
 with: \\
  $r_b(l_b,l)=[re_b(f_b)]^2$, and $f_b = l_b/l$. \\ \\
 We have now to take into account that the luminosity $l_b$ of the resolved star is not constant,
 but comes from the luminosity function.
 It was shown by Zhao (Zhao 1995) using the OGLE data that the number of resolved stars in
 Baade'swindow
 can be well described by a power law, with an exponent close to $-2$.
 To be precise we have a fraction of stars near luminosity $l_b$: 
 \begin{equation}
 \hspace{2.5cm} dF(l_b)= \beta_2 \ l_b^{-\alpha_2} \ l_0^{\beta_2}\ dl_b 
 \end{equation}
 with:
 $\alpha_2 \simeq 2$ and $\beta_2=\alpha_2-1$ \\ 
 consequently,
 $$
 d\tau(D_s,L,l_b)=d\tau(D_s,L) \times dF(l_b)
 $$

 To get the total optical depth associated with the unresolved stars, we have now to integrate over
 the luminosity range of these stars, from the absolute luminosity ($L_{min}$)  
 to the faint end of their luminosity function ($L_{end}$). 
 
 In addition, we have also to integrate over all the possible distances for the sources, and
 all apparent luminosities for the blends.\\ \\
 This leads to: 
 $$
 \tau= \int_{L=L_{end}}^{L=L_{min}}\int_{D_s} \int_{l_b=l_0}^{l_b=l_{sat}} \frac{dN(D_s,L)}{N_{tot}} 
 $$
 $$
 \times dF(l_b) \ R \ r_b(l_b,l) \ f(D_s)  \ dD_s \ dL \ dl_b
 $$
 \\
 If now we uses eqs, (1), (2), (3), we can express $\tau$ as:
 \\ \\
 $$
 \tau= \frac{R}{I_d} \times \beta_2 \ l_0^{\beta_2} \int_{D_s} \int_{L=L_{end}}^{L=L_{min}} \int_{l_b=l_0}^{l_b=l_{sat}}
 \ l_b^{-\alpha_2}  \ r_b(l_b,l) 
$$
$$
 \times \ lf(L) \ f(D_s) \ n(D_s) \ D_s^2   \ dL \ dl_b  \ dD_s 
$$
 This expression can be integrated over the variables $L$ and $l_b$,
 consquently:
$$
 \tau= \frac{1}{I_k}  \int_{D_s} \ N_L(D_s) \ f(D_s) \ n(D_s) \ D_s^{2-2\beta} \ dD_s 
$$
with:
$$
 N_L(D_s)=  \frac{I_k}{I_d}  \  R \ \beta_2 \ l_0^{\beta_2} \ D_s^{2 \beta}\int_{L=L_{end}}^{L=L_{min}} \int_{l_b=l_0}^{l_b=l_{sat}}
 \ l_b^{-\alpha_2}  \ r_b(l_b,l) 
$$
$$
 \times lf(L) \ dL \ dl_b 
$$ 
Note that the dependance in distance is hidden in the variable $l=L \times D_0^2/D_s^2$,
and also in the boundary $L_{min}$. \\ \\
Figure 5 illustrates the new $N_L(D_s)$ function for the unresolved stars in the case 
of the DUO and OGLE experiments. Note that these functiond reach their maxima at larger
 distances than the previous $N_{L0}(D_s)$ functions, it means that a large contribution
 will come from unresolved sources on the far side of the bar.
\begin{figure}
\centerline{\psfig{angle=0,figure=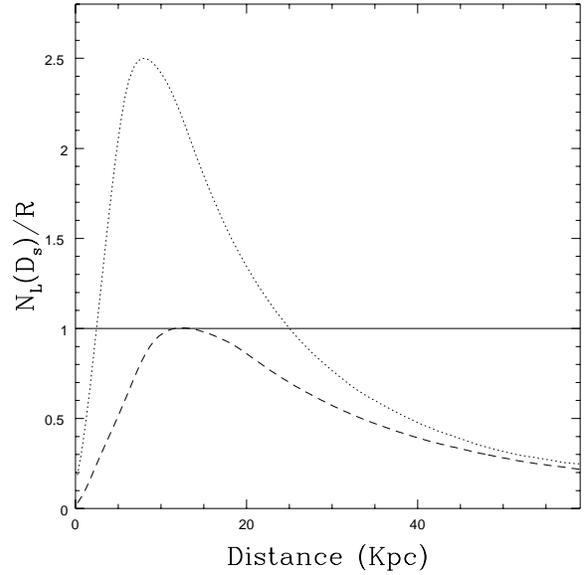,width=8cm}}
\caption{Plot of the $N_L(D_s)/R$ function as a function of distance for unresolved sources.
 The limiting magnitude defined for figure 1 were kept for this illustration.
  The dotted line is for DUO, and the dashed line for OGLE.} 
\end{figure}
 It is now possible to quantify the contribution of the unresolved sources to the total optical
 depth. In table 2 the ratio of the optical depth of  unresolved sources to the the optical depth
 of resolved sources 
 is indicated for DUO and OGLE. The calculation of this ratio require an estimation of the
 R constant. A crude estimate of R is the ratio of the surface covered by the resolution radius
 to the mean area occupied by a star. For the DUO experiment the resolution radius is close to
 3 pixels, and the mean area occupied by a star is about 60 pixels (Alard and Guibert 1996).
 This gives $R \simeq 0.5$. For OGLE the resolution radius is probably close to the seeing
 value (this is due to smaller pixels). Consequently with a mean seeing of $1.25 \arcsec$ in
 Las Campanas, and a pixel of $0.44 \arcsec$ we can estimate the resolution radius as 2.7
 pixels. The OGLE experiment follows $1.3 \ 10^6$ stars on 14 images of $2048 \times 2048$
 pixels each. This gives again a value for R, of $R \simeq 0.5$.
\begin{table} 
   \caption{Ratio of the optical depth of  unresolved sources to the the optical depth
     of resolved sources .
    A value of $R=0.5$ is adopted for the calculations (see text for discussions).}
   \begin{tabular}{llll}
      \hline\noalign{\smallskip}
    $ \tau/\tau_0$ & Bulge lenses & Disk lenses & Total \\ 
   \hline\noalign{\smallskip}
     OGLE & 0.58 & 0.55 & 0.57 \\ 
     DUO &  1.93 & 1.81 &  1.89\\  
   \hline\noalign{\smallskip}
      \noalign{\smallskip} 
  \end{tabular}
\end{table}
%
%
%%%%%%%%%%%%%%%%%%%%%%%%%%%%%%%%%%%%%%%%%%%%%%%%%%%%%%%%%%%%%%%%%%%%%%%%
                   \subsection{The rates of microlensing events from unresolved stars.}
%%%%%%%%%%%%%%%%%%%%%%%%%%%%%%%%%%%%%%%%%%%%%%%%%%%%%%%%%%%%%%%%%%%%%%%%
 The rates per resolved star for unresolved events can be calculated in the
 same way as for the Optical Depth. \\
 We have:
\begin{equation}
 \Gamma =  \frac{1}{I_k} \int_{D_s} \ N_{L\gamma}(D_s) \ f_{\gamma}(D_s) \ n(D_s) \ D_s^{2-2 \beta} \ dD_s 
\end{equation}
with:
$$
N_{L\gamma}(D_s)=  \frac{I_k}{I_d}  \  R \ \beta_2 \ l_0^{\beta_2} \ D_s^{2 \beta} \int_{L=L_{end}}^{L=L_{min}} \int_{l_b=l_0}^{l_b=l_{sat}}
 \ l_b^{-\alpha_2}  \ re_b(l_b,l) 
$$
$$
 \times lf(L) \ dL \ dl_b 
$$ 
It is clear that the change of the Einsten radius by the factor $re_b(l_b,l)$ changes also
 the duration by the same factor. Consequently in the differential rates calculation, each duration
 $t_E=r_E/V$ will have to be scaled by the factor $re_b(l_b,l)$. However, we see in equation 4
 that for a given Einsten radius there is a distribution of the scaling factors $re_b(l_b,l)$
 hidden in the double integral $N_{L\gamma}(D_s)$. This distribution can be calculated for
 each distance $D_s$, and for the realisation of the calculations,
 the scaling factor of the Einsten radius will be choosen by a Monte-Carlo
 method using this distribution. Some examples of such distributions for different distances are 	
 shown in figure 6, in the case of the DUO experiment. \\ \\
\begin{figure}
\centerline{\psfig{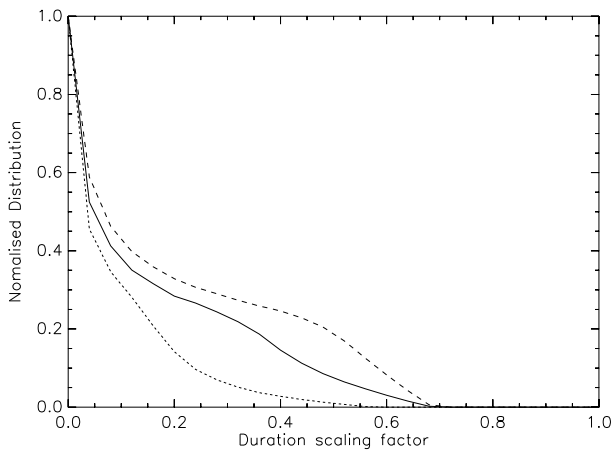}}
\caption{Examples of duration scaling factor distribution functions, for different distances.
 The solid line is for a distance of 8 Kpc, the long dashed line is for 15 Kpc, and
 the short dashed line is for 6 Kpc.} 
\end{figure}
 The result of the calculation of the $\Gamma_0$ and $\Gamma_k$ rates is shown in figure 7 for
 DUO, and in figure 8 for OGLE. It is evident that the DUO experiment is dominated
 by unresolved stars. This result is in very good agreement with the data 
  (Alard and Guibert 1996). Even for OGLE which has the deepest photometry, the bias is
 still very significant. It is rather straightforward to relate these events to
 the excess in the rates and optical depth observed towards the Bulge. This excess was not
 been explained, even using a bar in the Galactic model. We find here a natural explaination.
\begin{figure}
\centerline{\psfig{angle=0,figure=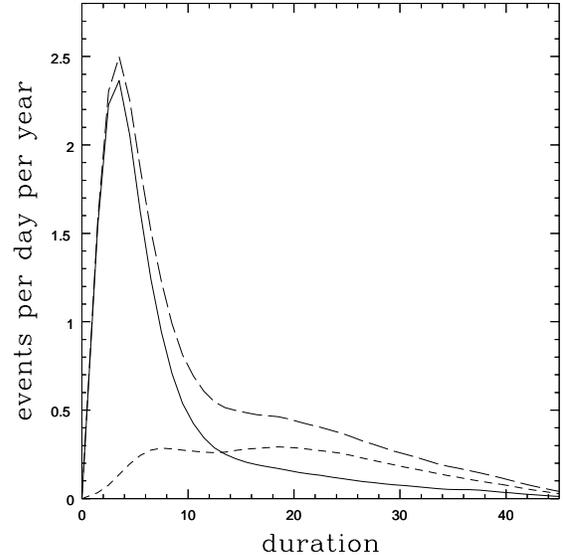,width=8cm}}
\caption{Comparison of the the contribution to Bulge microlensing event rates
 for resolved stars (short dashed line), and unresolved stars (solid line). 
 The total rate is represented by a long dashed line.
 The  rates are estimated for the 1994 season of the DUO experiment using the DUO efficiencies
 (Alard and Guibert 1996). Note the dominant contribution of the unresolved stars.} 
\end{figure}
\begin{figure}
\centerline{\psfig{angle=0,figure=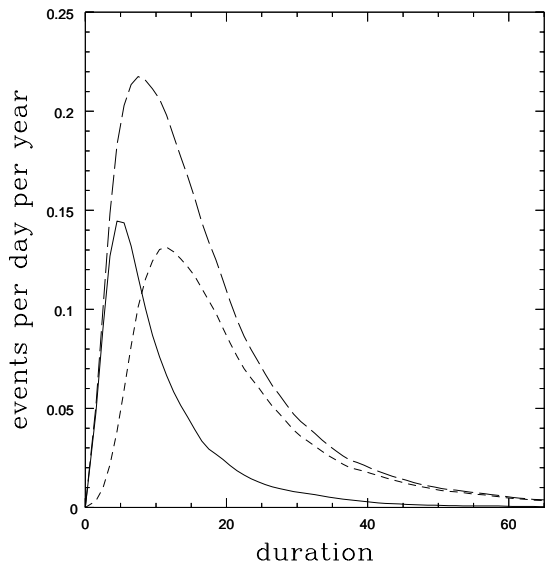,width=8cm}}
\caption{Comparison of the contribution to Bulge microlensing event rates
 for resolved stars (short dashed line), and unresolved stars (solid line). 
 The total rate is represented by a long dashed line.
 The rates
 are estimated for $10^6$ stars in Baade window using the OGLE efficiency
 (Udalsky {\it et al.} 1994).} 
\end{figure}
%%%%%%%%%%%%%%%%%%%%%%%%%%%%%%%%%%%%%%%%%%%%%%%%%%%%%%%%%%%%%%%%%%%%%%%%%%%%%%%%%%%%%%%%%%
%%%%%%%%%%%%%%%%%%%%%%%%%%%%%%%%%%%%%%%%%%%%%%%%%%%%%%%%%%%%%%%%%%%%%%%%%%%%%%%%%%%%%%%%%%
%
\section{Can we identify the blending events using the light curves ?}
 The light curve of a blended event is affected by the additional light
 coming from the companion, so that the light curve is modified.
 This blended light curve has more extended wings than an unblended event,so that
 it might be possible
 to differentiate it from the unblended one using a chi-square test. \\
 Another important issue is the possibilty that the amplified star and
 its companion may have different colors. For instance if the companion is more
 blue than the amplified star, we expect more blending in the blue, and thus
 a larger reduction of the amplitude in this color than in the red. 
 
\subsection{Color variations}
 Let us now investigate the problem of color variation during the event in more detail.
 In the current microlensening experiments the sources are concentated in the range $18<V<20$.
 An examination of Terndrup (Terndrup, 1988) color magnitude diagram in Baade'swindow indicates\
 that the mean
 $V-I$ color is about 1.3 to 1.4. For fainter stars, where expect to 
 find the unresolved sources, the Holtzman {\it et al.} color magnitude diagram shows that 
 the color changes again very slightly. The mean color a close to 1.4 at $V=22$.
 We see that in the apparent magnitude range of interest we expect a diferential color variation 
 of about 0.1 magnitude at most between 
 the unresolved source and the blending companion. we have to reach $V=23$ to expect
 more significant color variation between sources and blend (about 0.2 magnitudes). 
 Unfortunately the photographic technique does not perform very well in the red band,
 thus an accurate investigation of the color changes during the event is not possible.
 It means that the DUO survey will not be able to detect the slight
 color changes expected. The OGLE survey has a good coverage only in the I band, the
 sampling in V is very sparse, and does not allow color analysis during the event.
 In the MACHO case, photometric data are available in two large bandpasses, which is transformed
 to a $V-R$ color index. The analysis of the LMC events by MACHO shows that they can
 identify color variations due to blending for some of the events. However there is
 a small residual noise on the color variations of all the events of about 0.02 to 0.03 magnitudes,
 associated with the photometric errors. At a 3 $\sigma$ confidence level we may require
  a color variation of about $0.075$ magnitude to firmly established the chromaticity of 
 the event. Converted to $V-I$ it requires a color changes of about 0.15 magnitudes at least
 to recognise a blended event with a good confidence level. \\ 
 We consider now with a such accuracy how many of the Bulge microlensing events which involves
  unresolved
 sourves could be identified as such. \\ \\
 Let us make the following simple model: \\ \\
 And we assume that the difference of color can be described by a gaussian distribution,
 shifted by a systematic value. Thus the number density $n(C)$ of stars with color a difference in
 color C will be expressed as:
 \begin{equation}
 n(C)=\exp \left ( \frac{(C-{\rm shift_c})^2}{-2 \ \sigma_c^2} \right )
 \end{equation}
 The Holtzman {\it et al.} color magnitude diagram
 indicates that for the gaussian we can assume $\sigma_c \simeq 0.07$. In the range $V=20$ to
 $V=22$, a conservative value for the shift is ${\rm shift_c} \simeq 0.15$.
 The color shift is then given by:
 $$
 \Delta_C=2.5 \log(A_i/A_v)
 $$
 i and v stand where the subscripts for the two photometric colors.
 with:
 $$
 A_i=\frac{1+a f_b}{1+f_b}
 $$
 and:
 $$ 
 A_v=\frac{1+a f_b c}{1+f_b c}
 $$
 \\ 
 a and c are defined by the following expressions: \\ \\
 $$
 a = \frac{u^2+2}{u \times \sqrt{u^2+4}} \hspace{0.75cm} c=10^{ \ C/2.5}. 
 $$
 The number of events with color blending signature identified will be simply the number
 of events with a color shift $\Delta_C>0.15$. It can be easily calculated if we assume that the unblended
 impact parameter $u$ has a uniform distribution. In this case it is sufficient to integrate the
 number of events with $\Delta_C>0.15$ over the $u$ distribution with a given color difference
 $C$. The calculation is completed by suming over the color difference distribution
 defined in eq (5).The maximum value of the impact parameter is set
 by the blending factor $fb$ using the formulae:
 \begin{equation}
  u_{max}=\left [2(-1+A_b/\sqrt{(A_b^2-1.0)} \right ]^{1/2}
 \end{equation}
 with:
 $$
 A_b = 0.34 \times f_b+1.34
 $$
The calculation of the percentage of events with color shifts is performed for several values
 of the blending factor $f_b$. An illustration of these calculations is given in Figure 9,
 where the color shift is expressed as a function of the impact parameter for a few $f_b$ values,
 and color diffrence of 0.2 magnitude. Looking at this diagram we can already guess that the
 number of events with detectable color shitfts will be small. The final values of the events with
 color shifts identified is given in table 3. The values are extremly small. A shift
 of 0.15 mags in the color difference distribution, even for high values of the blending factor. 
  The values for a shift of 0.3 magnitudes are also given, to illustrate the case of the very faint
 unresolved stars. However these stars will be extremly blended, and consequently will give
 events will very short duration which will be almost all removed by the low efficiency of the
 experiments in this range. Consequently the color shift method will give poor results in the case of the Galactic
 Bulge. In the case of the Magellanic Clouds, the color changes rapidly with the magnitude close
 to the limit of the microlensing experiments, which explains why so many events with color shifts
 are found by MACHO (Alcock, {\it et al.}, 1996)towards the LMC.
\begin{figure}
\centerline{\psfig{angle=0,figure=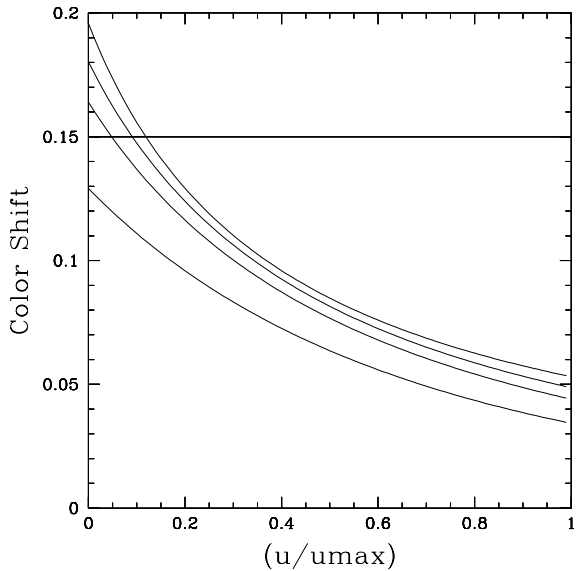,width=8cm}}
\caption{The distribution of the color shift for blending parameters of 2,5,10,50, and
 a color difference of 0.2. The curve with $f_b=50$ is upper, and the values 10,5,2, 
 are below in a decreasing order.} 
\end{figure}
\begin{table} 
   \caption{The fraction of events with detectable color shifts.}
   \begin{tabular}{lllll}
      \hline\noalign{\smallskip}
     Blending factor &2 & 5 & 10 & 50 \\ 
   \hline\noalign{\smallskip}
      ${\rm shift_c}=0.15$ & 0.0025  & 0.0219 & 0.0364 & 0.0485 \\ 
      ${\rm shift_c}=0.3$ & 0.1677  & 0.3079 & 0.3559 & 0.3906 \\
   \hline\noalign{\smallskip}
      \noalign{\smallskip} 
  \end{tabular}
\end{table}
\subsection{The shape of the light curves.} 
 I will investigate another possibility to identify a blended microlensing event
 based on the shape of the light curve in this
 section. The blending of light modifies the shape of the light curve, and consequently
 we may use a statistical test to see if a light curves differs significantly from
 an unblended one. However we have to realise that the likely difference due
 to blending is rather small, and we may expect that with the current available 
 photometry it will be difficult to identify a blended light curve at a significant 
 level of confidence. It is possible to adress this problem in more general terms
 using Monte-Carlo simulation of microlensing events. Assuming a noise distribution,
 it is easy to simulate microlensing events with different blending factors. These blended
 events can be analysed by fitting an unblended curve to the simulated data set. We expect
 a systematic difference in the chi-square of the fir due to the different shape of the blended light
 curve, especially in the wings. However the problem is to know how significant is this difference
 compared to the normal chi-square variations for unblended events due to noise.
 This problem can be easily addressed if we are able to build the chi-square distribution for a series
 of blended light curves. Such chi-square distributions are illustrated in Fig. 10,
 the line shows the limit within which 95 percent of the chi-square distribution for unblended events is contained. Beyond
 this limit we have a 95 percent confidence level that the chi-square indicates a light curve
 which is systematically different from the unblended model. 
\begin{figure}
\centerline{\psfig{angle=0,figure=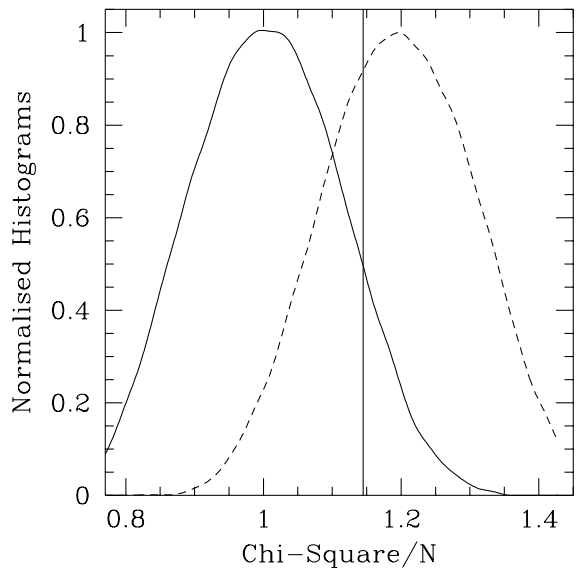,width=8cm}}
\caption{An example of simulated Chi-square distributions for unblended events (full line),
 and for blended events (dashed line). The vertical line indicates the 95 percent limit for the
 unblended distribution (see text for explanations). For this example the events are simulated with
 a duration of 40 days, and an impact parameter of 0.05. The blending factor has a value of 4.} 
\end{figure}
It is now sufficient to sum the fraction
 of the blended distribution beyond this limit to get the fraction of blended events $R_b$ which can be
 identified with a good confidence level. The ability to recognise a blended event will of course
 depend on the amplitude of the event, for a given blending factor $fb$ and a given duration $t_0$. 
 The amplitude is related to the impact parameter $u$. Consequently to get the expected fraction $R_b$
 , at $fb$ and $t_0$ it is sufficient to sum on the impact parameter $u$, in the range 0 to $u_{max}$ (
 $u_{max}$ is defined in eq 6). 
 The result of the corresponding calculation is illustrated in Fig 11,
 as a function of the blending factor, for a series of durations. The blended events are well identified only
 for small impact parameters. Beyond $u_{max}/10$, the efficiency drops significantly, which makes the
 total efficiency quite low.
\begin{figure}
\centerline{\psfig{angle=0,figure=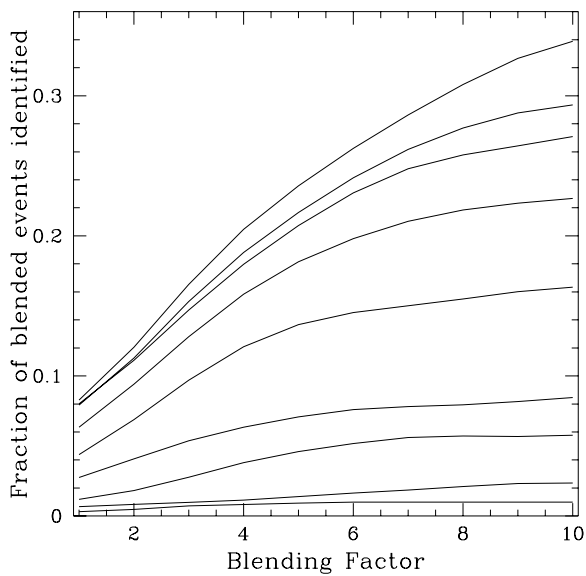,width=8cm}}
\caption{The fraction of blended events identified as a function of the blending factor for
 different durations. The durations are respectively 
100,63,25,15,10,6,4,2.5 days (from top to bottom).} 
\end{figure}
 These tabulated expressions of $R_b$ as a function of $fb$ and $t_0$ are now introduced directly
 in the calculation of the unresolved event rates. Then new rates corrected for the number
 of blended events which could be recognised are computed. Fig 12 shows a comparison between this
 rate and the uncorrected rate in a case resembling to the OGLE experiment. Typical errors distributions
 are extrpolated from the DUO experiment (Alard and Guibert 1996), and are divided by a scaling 
 factor of 2 to take into account the better quality of the OGLE photometry. The resulting errors
 distribution has a sigma about 0.07 magnitude for most of the points, and reaches 0.1 or slightly more
 in 10 percent of the cases; it seems to be an acceptable description of the OGLE photometry on
 stars close to the limiting magnitude, which represents most of the sample.
 We see that the difference
 is small, and considering that OGLE has the best photometric accuracy, it is certainly worse
 again for the others experiments. \\ \\
 The conclusion is again, as for the color test, that given the mean quality of the 
 photometry available in the microlensing experiments only, a slight percentage of blended
 events should be recognised on the basis of the light curve shape. Only the the events with good
 photometry and small impact parameter should be identified as blended.
 However for some of the monitored
 stars the OGLE experiment is able to perform photometry much better again than for the majority
 of the stars. This gives an interesting opportunity to look for a blending signature in the light curve.
\begin{figure}
\centerline{\psfig{angle=0,figure=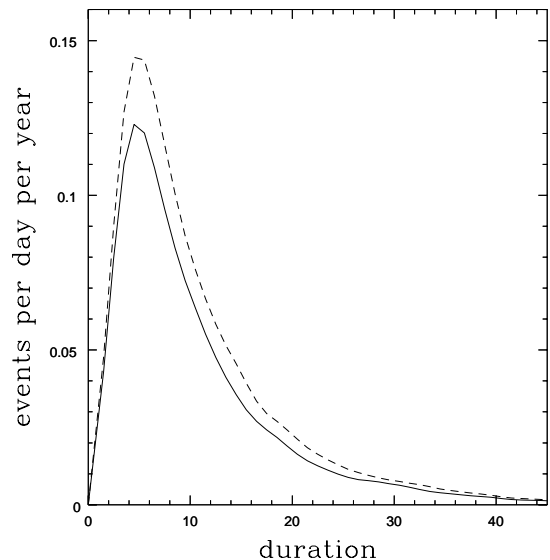,width=8cm}}
\caption{The event rate statiscally corrected for the fraction of blended which
 might be identified on the basis of their light curve shape. The corrected rate is represented
 by a full line, and the initial rate by a dashed line.} 
\end{figure}
\section{OGLE 5: lensing of an unresolved star} 
 The photometry achieved on the OGLE 5 event is of an interesting quality, the errors bars are as good
 as a few percent on many points. However this event is not well fitted with the standard unblended
 model (see figure 13). The discrepancy is especially large in the wings of the event, so that it is
natural to try to fit this event with blending. Fig. 14 shows the dramatic improvement of the
chi-square when a blended light curve is fitted. The best fit is obtained for a blending factor of
 2.45 (Fig 15.). Thus makes the lensed source 1.34 magnitude fainter, and places it below the OGLE detection
 limit. Consequently it is very likely that OGLE 5 is an example of lensing on an unresolved star.
 The better Chi-Square per degree of freedom is different from unity, but we have to recall that
 an event on an unresolved star is seen only by blending on a resolved star. For the photometry,
 the position of the resolved star is used, but it might be significantly different from the
 position of the true magnified star. The achieved errors in a PSF fitting routine are rather 
 sensitive to the quality of the positioning, thus we expect somewhat larger errors in the case
 of an unresolved event. The small discrepancy in the Chi-Square per degree of freedom is thus
 perfectly consistent with the scenario of an unresolved event.
\begin{figure}
\centerline{\psfig{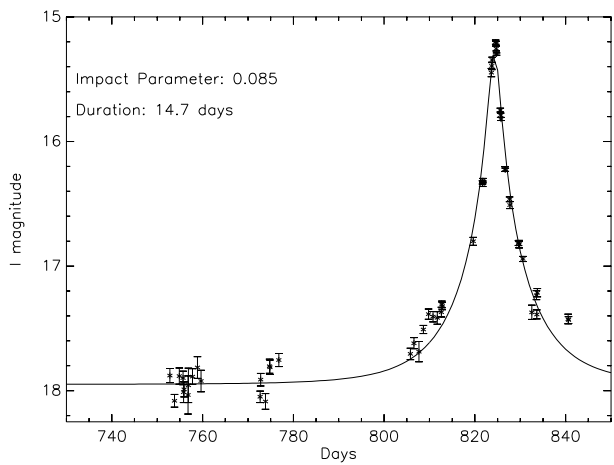}}
\caption{The fit of an unblended light curve to the data, note the large discrepancy
 in the wings.} 
\end{figure}
\begin{figure}
\centerline{\psfig{angle=0,figure=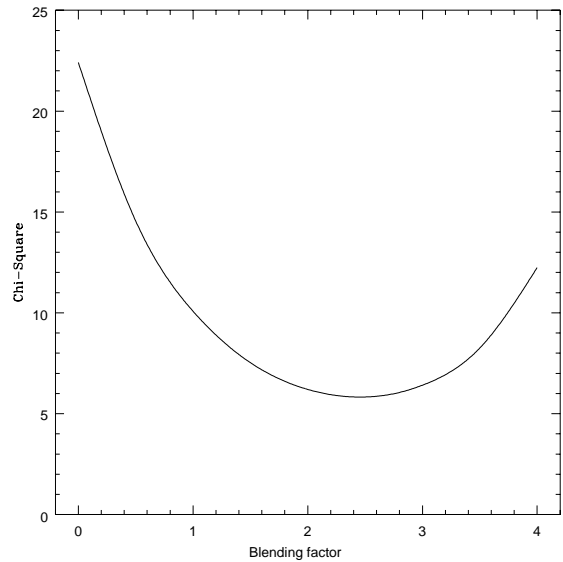,width=8cm}}
\caption{The Chi-Square per degree of freedom obtained for a fit to the data for OGLE 5 with
 different blending factors.} 
\end{figure}
\begin{figure}
\centerline{\psfig{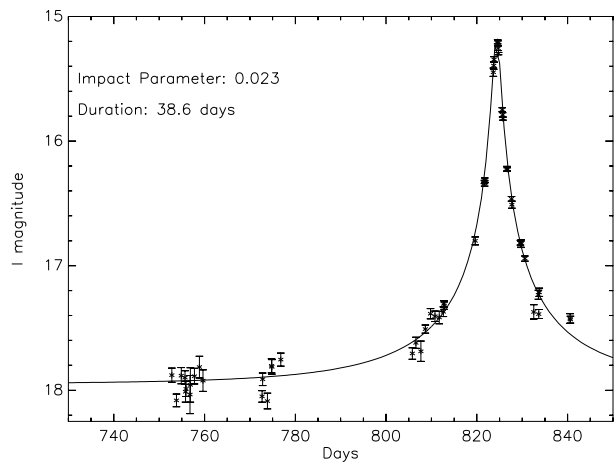}}
\caption{The best fit with blending, note the large change in the estimated 
 duration in comparison with the unblended fit.} 
\end{figure}
\section{Discussion}
In this paper I show that the rates of microlensing events and optical depth to
 microlensing due to the
unresolved stars can be modelled. However I emphasize that the exact
 rates and optical depth are closely related to the number of
 unresolved stars which will contribute per resolved star. This number
 is well constrained in crowded fields, essentially because the surface
 occupied by a star on the image is set by the crowding limit. On the
 other hand, this crowding limit is closely related to the resolution
 radius, thus the surface occupied by a star is just a function
 of this resolution radius. A value of 0.5 was found for $R$ (the ratio
 of the surface covered by the resolution radius to the mean
 surface occupied by a star), both for DUO and OGLE. This simply means
 that probably there is a linear relation between the resolution radius
 and the mean radius per star. \\
 However in the crowded fields, if a new star appear at more than a resolution
 radius, it does not mean necessarily that it will be resolved. There are so
many stars that it might well be confused with another one. This issue
 should be clarified later using Monte-Carlo simulation of crowded fields.
 This effect will probably lead to a slight increase of $R$.
 Another problem is that if the amplified star is situated at some distance 
 from the resolved star the photometry which assumes fixed positions (except
 for DUO) is certainly affected. This possibilty was already discussed in the case
 of OGLE 5 and will cause a slight drop in the efficiency, which will increase
 as the unresolved star will be more distant form the resolved one. 
 This issue
 could be also clarified later using Monte-Carlo simulation.
 But globally,
 this effect will more or less cancell out with the confusion effect exposed just
 before. Thus to conclude, the value $R=0.5$ might be an acceptable approximation.
\begin{figure}
\centerline{\psfig{angle=0,figure=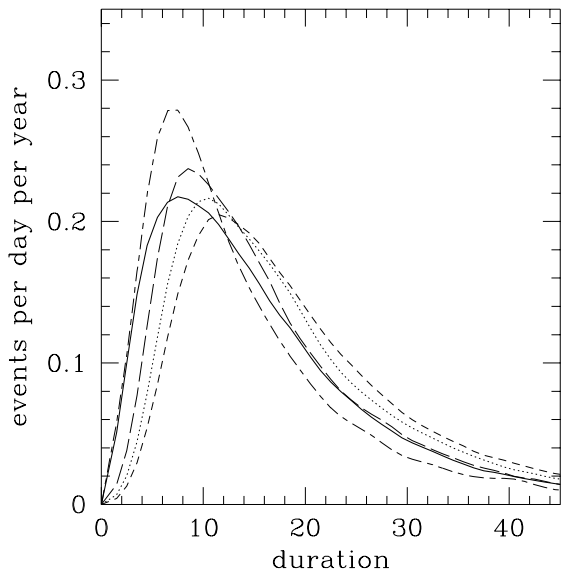,width=8cm}}
\caption{ Comparison of the total microlensing rates (unresolved+resolved stars) with the rates for 
 resolved stars for different cut-offs in the mass function 
 in case of the OGLE experiment. The cut-offs are: 0.02 $M\sun$ (long and short dashed line), 0.04 $M\sun$
 (long dashed line), 0.06 $M\sun$ (dotted line), and 0.08 $M\sun$ (short dashed line).
 The scaling factors are respectively: 1.09, 1.24, 1.18, 1.28 (see text for explainations).} 
\end{figure}
\section{Conclusion.}
The modeling of the Unresolved star microlensing rates demonstrates that they induce an important
 bias for the short events. I show that given the current photometric errors, most
 of these events would be hard to distinguish from the unblended events. Thus
 this is an important and annoying bias. If this bias is not taken into account,
 it will certainly influence the shape of the lens mass function estimated
 from the data. For instance, if a Salpeter mass function is used, we expect
 that the lower cut-off will be shifted towards the brown dwarfs. This idea is
 illustrated in Fig. 16 where the total OGLE rates (resolved+unresolved) are
 compared to the rates for Salpeter mass functions with different lower cut-offs 
 (I recall that
 initially I used a Salpeter mass function with a lower cut-off of $0.08 M\sun$ to compute the OGLE and DUO
rates). 
A better agreement with the total rates is found if the different trial
 models are slightly scaled. I calculated this scaling factor, so that the
 trial models give the same total rates in the range 0 to 80 days as the
 total rate (see Fig 16.). The case of the DUO experiment will be treated 
 in another paper (Alard and Guibert 1996). 
 While the unresolved star bias has to be taken into account, as demonstrated
 in the previous section, the modelling includes some uncertainties. 
 However for the data set already assembled by the microlensing experiments, it is
 certainly an acceptable solution to the unresolved star bias. For the future, a very 
 interesting possibilty is offered by the PLANET (Sackett, 1995) collaboration, who will provide
  a dense and accurate sampling of the events. It will allow a much better
 control of the bias caused by the unresolved, using both color shift but importantly the shape
 of the light curve, as demonstrated for OGLE 5. \\
 To conclude, I will also emphasize that the contribution of the unresolved stars
 drops if the magnitude limit is increased. This is well illustrated by the comparison
 OGLE vs. DUO. Consequently a first solution is to try to reach stars as faint as possible
 to minimise the bias. It is certainly a possible orientation for the OGLEII experiment,
 which will achieve improved resolution better again with a new telescope. On the other hand
 for MACHO and EROSII experiments, which use bigger pixels, and cover larger fields,
 an interesting solution is certainly to monitor the bright clump giants. These stars are
 much brighter (about 3 magnitudes) than most of the bulge stars, and their density per
 image is also much lower. Thus, the unresolved stars per giant will be low and they 
 will be so blended  that the effective Einsten radius will be very small. This idea
 leads to the conclusion that the  unresolved star bias will be very small on clump giants.
 Some other advantages of the clump giants include the good photometry expected, and
 also the possibility to have a sample with a very high completeness (Gould 1995).
\begin{acknowledgements}
It is a great pleasure to thank B. Paczy\'nski, Gerry Gilmore, Olivier Bienaym\'e and
 Gary Mamon for interesting discussions.
\end{acknowledgements}
\end{document}